\documentclass[conference,10pt]{IEEEtran}
\IEEEoverridecommandlockouts
\usepackage{cite}
\usepackage{amsmath,amssymb,amsfonts}
\usepackage{algorithmic}
\usepackage{graphicx}
\usepackage{textcomp}
\usepackage{xcolor}
\usepackage{multirow}
\usepackage{multicol}
\def\BibTeX{{\rm B\kern-.05em{\sc i\kern-.025em b}\kern-.08em
    T\kern-.1667em\lower.7ex\hbox{E}\kern-.125emX}}
\begin{document}

\title{Approximate Scan Flip-flop to Reduce Functional Path Delay and Power Consumption}

\author{\IEEEauthorblockN{Lakshmi Bhanuprakash Reddy Konduru}
\IEEEauthorblockA{\textit{Dept. of Electrical Engineering} \\
\textit{IIT Tirupati}, Tirupati, India\\
ee19d501@iittp.ac.in}
\and
\IEEEauthorblockN{Vijaya Lakshmi}
\IEEEauthorblockA{\textit{Dept. of Electrical Engineering} \\
\textit{IIT Tirupati}, Tirupati, India \\
ee20d001@iittp.ac.in}
\and
\IEEEauthorblockN{Jaynarayan T Tudu
}
\IEEEauthorblockA{\textit{Dept. of Computer Science Engineering} \\
\textit{IIT Tirupati}, Tirupati, India \\
jtt@iittp.ac.in}
}
\maketitle
\begin{abstract} \footnote{This work has been partially funded by Science and Engineering Research Board, Govt of India, under project grant EEQ/2019/000714.}

The scan-based testing has been widely used as a Design-for-Test (DfT) mechanism for most recent designs. It has gained importance not only in manufacturing testing but also in online testing and debugging. However, the multiplexer-based scan flip-flop, which is the basic building block of scan chain, is troubled with a set of issues such as mux-induced additional delay and test power among others. The effect of additional delay due to the multiplexer on the functional path ($D_{in}$ path) has started influencing the clock period, particularly at the lower technology nodes for the high-performance design. In this work, we propose two scan flip-flop designs using $10nm$ FinFET technology to address the problem of mux-induced delay and internal power. The proposed designs have been experimentally validated for performance gain and power reduction and compared to the existing designs.
\end{abstract}


\section{Introduction}
Testing the complex SoC designs has been a challenging task over the years. As the complexity of the design increase, the test challenges with respect to test power and test time have been growing. The challenges are due to the scan-based Design-for-Test (DfT) architecture which has been used in most contemporary designs as a trustworthy DfT solution. The scan-based DfT has become popular due to its simplicity and useful features such as debugging capability. 
Back to back connected scan flip-flops forms scan chain and is the basis for scan-based DfT architecture.
A master-slave D flip-flop is converted to a scan flip-flop by adding a multiplexer to switch between the functional mode and test mode. Multiple such flip-flops form a scan chain, as shown in Figure \ref{fig:fullscan}. The data in (DI), scan in (SI), scan enable (SE), clock (CLK) are the inputs to scan chain and scan out (SO) being the output. In presence of the clock signal, the scan chain is operated to load, apply, and unload the necessary patterns. The test patterns are shift-in to load into the scan chain and then launched to sensitize and propagate the fault effect to outputs and pseudo-outputs. The response is captured back and then unloaded to observe at the primary output or compacted at the MISR (multiple input signature register).

The test patterns which are being shifted through the scan chain create excessively higher switching activity in the circuit, which causes high dynamic power dissipation. It has been observed in several experiments that the test power is relatively very high compared to the normal functional power of the circuit \cite{Girard2002}. There are several components of power dissipation. The dynamic power dissipation which is being caused by the switching activity has been a cause of concern for the contemporary designs at lower technology nodes at below $32~nm$. The switching activity could either come from the internal or external switching of a given design. For example, the internal switching activity of a 2-input NAND gate would differ for different inputs, whereas the external switching activity, also known as output switching activity, may remain the same. Similarly, the internal and external switching activity also differs for a flip-flop. As the number of flip-flops in a design increases, the internal switching begins to dominate the total power dissipation. The problem of test power, test time, and test data volume has been there since the advent of scan architecture in the early 80s. The scenario of the problems, however, has been changing as the complexity, clock frequency, and transistor size change. We will be addressing two problems in this work:
\begin{itemize}
\item the problem of additional delay in the functional path due to scan flip-flop and
\item the excessive power dissipation due to internal switching in a scan flip-flop.
\end{itemize}

The additional functional path delay comes from the multiplexer present in the scan flip-flop. This raises a concern for a high-performance digital design where every small unit of time plays a major role in the overall performance of the design. Therefore, the extra delay due to the scan DFT is unacceptable, but it has become unavoidable. We address this problem by designing a new scan flip-flop that eliminates the gate delay that is due to the multiplexer. The second problem is rather severe as it can potentially cause chip burn-out due to excessive heat dissipation during testing. The switching power has become again a serious problem even though the technology has moved to FinFET. The proposed flip-flop design also addresses this problem, particularly the internal switching power.


The problem of additional multiplexer delay has been addressed in the earlier work by Ahlawat et al \cite{ahlawat2018high, ahlawat2015new} among others. However, the designs or techniques proposed in the previous works have an area and power overhead. The work \cite{naeini2015novel} uses gating techniques to reduce the test power.
The low power test methods are categorized as ATPG-based, DfT-based, and system-level methods\cite{Nicolici2007}, \cite{Girard2002}. The ATPG-based techniques target the scan shift power, test pattern reordering, and reduce the transition test vectors \cite{Remersaro2006}, \cite{Dutta2013}. This method can reduce the redundant combinational switching but not entirely. In \cite{ahlawat2016minimization}, the authors used the test vector reordering method to minimize the intra-pattern switching. DFT-based techniques include scan chain partitioning, reordering, power supply gating, and output gating. The works \cite{Tuduscangraph2010,Tuduscanreoder2010,TuduScan2010} describe the low-power scan shift methods based on the scan chain reordering. In \cite{Gerstenminimized1999}, the authors proposed a blocking logic by using NAND or NOR gates as blocking elements and freezing the combinational path inputs to either logic 1 or logic 0 during the scan shift mode. Still, in the functional mode, it deteriorates the timing performance of the circuit. In \cite{Devanathanpmscan2007}, the authors use a transmission gate as a blocking element to block the scan ripple propagation. A pull-up or pull-down logic keeps the combinational inputs at constant logic 1 or logic 0 values. Also, the blocking logic consumes a significant amount of power in functional mode. The other techniques to design the low-power scan flip-flop include \cite{Seoscan2015,zhang2014low,tekumalla2013low}. In \cite{mishra2010modified}, a method to isolate the slave latch during scan shift operation is proposed by introducing an extra transmission gate. A dynamic slave latch is used to propagate the scan outputs. This method reduces the combinational switching activities at the cost of increased area and performance overhead.

This work proposes a new scan flip-flop design that consumes low power and occupies a low area. The proposed method removes the transmission gate present in the functional path and the primary input directly connected to the scan path output, thereby reducing the delay compared to the multiplexer-based design. In the other design, we replaced the multiplexer with the GDI-based technique.

The rest of the paper is organized as follows: Section
II describes the background and different modes of operation of the scan flip-flops. Section
III is dedicated to different types of the proposed scan flip-flops design. We discuss the experimental results and comparison with the existing designs in Section IV. Finally, the paper is concluded in section V.
\section{Background}
\begin{figure}[!htbp]
    \centering
    \includegraphics[scale=0.17]{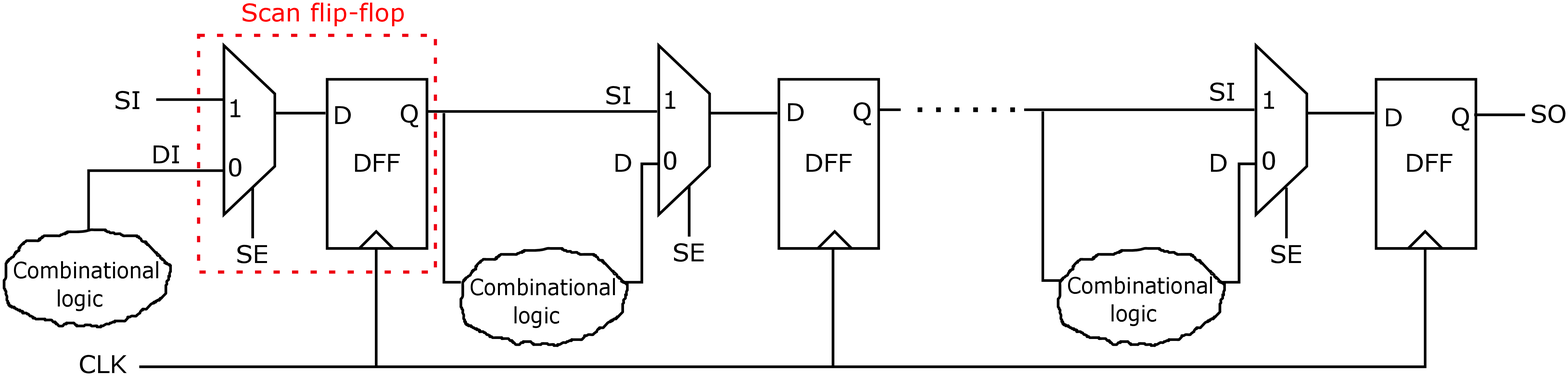}
    \caption{General full scan architecture. DFF is the master-slave D flip-flop, CLK = clock, SI = scan input, DI = data input, SE = scan enable and Q = output.}
    \label{fig:fullscan}
\end{figure}

In the full scan architecture (as shown in Fig. \ref{fig:fullscan}), all the conventional flip-flops are replaced by scan flip-flops that operate in functional and test modes depending on the scan enable (SE) control signal. During functional mode, SE is made low, and the circuit performs normal operation where the flip-flops get inputs from the combinational part. Whereas in the test mode, SE is made high, and all the flip-flops are connected serially to form a shift register structure.

\subsection{Functional mode}
In this mode, the scan enables the signal to be set to `0' so that the scan input signal is disconnected. All the flip-flops get inputs from the combinational logic, and the design works regularly. The only change is an additional delay is introduced due to the multiplexer in the input path. For the scan flip-flop design in Fig. \ref{fig:APPROXSFF}, the transistors P1, P2, and M1-3 form the master part, and P3-5, M4, and M5 transistors form the slave part of the D flip-flop. Master and slave are functional during the positive and negative edges of the clock, respectively. The D flip-flop is a negative edge-triggered flip-flop. The same D flip-flop design is used in all the proposed scan flip-flop implementations, and they differ in the multiplexer part of the scan flip-flop.
\subsection{Test mode}
\begin{figure}[!htbp]
    \centering
    \includegraphics[scale=0.21]{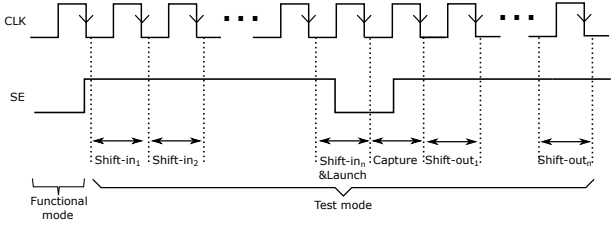}
    \caption{Timing Diagram}
    \label{fig:timedia}
\end{figure}

During this mode, the scan enable signal is set to `1' (except for the capture step) so that the data input is disconnected. All the flip-flops in the design form a shift register structure (also called a scan chain) where the output of one flip-flop drives the next subsequent flip-flop. The test vector is applied at the scan input. There are three steps in the test mode: shift-in, capture, and shift-out.\\
\textit{Shift-in:}
The test input bits are loaded into all the flip-flops at this stage. The number of clock cycles required to completely load the test inputs into all the flip-flops is the same as the number of flip-flops (length of scan chain).\\
\textit{Shift \& Launch:}
The last cycle during the shift-in operation is called shift-in and launch, which is the combined operation of shift-in and launch where the outputs of all scan flip-flops get applied to the combinational circuit. The scan enable signal is made low during this cycle.\\
\textit{Capture:}
This step requires one clock cycle to capture all the test bits loaded in the shift-in step into the flip-flops. The scan enable signal is set to `0' during this step.\\
\textit{Shift-out:}
This is the unloading step where the latched values in all the flip-flops are transferred to the output serially. This step requires the number of clock cycles same as in the shift-in step. The scan enable signal  is set back to `0' during this step.\\
\textit{Timing diagram:}
Fig. \ref{fig:timedia} shows the different modes of a scan flip-flop achieved by varying SE signals. Assuming the scan chain has n flip-flops, shift-in operation during test mode requires n clock cycles. The last shift-in cycle is the combined operation of shift-in and launch, where the outputs of all scan flip-flops get applied to the combinational circuit. The SE is made high during the shift-in operation except in the last shift-in cycle, where SE is made low.  The shift-in is followed by a capture cycle where SE is kept low so that all the flip-flops get inputs from the combinational circuit. Next comes the shift-out operation, where the scan output is received serially. It takes n clock cycles for the shift-out, and SE is made high during the operation.
\\
The minimum clock period, $t_{CLK} \ge t_{cq} + t_{comb} + t_{su}$\\
$t_{comb}$ = combinational path delay,\\
$t_{su}$ = setup time,\\
$t_{cq}$ = clock to q delay\\
The maximum clock frequency, $f_{CLK} = \frac{1}{t_{CLK}}$

\section{Proposed Scan Flip-flops}

\subsection{Approximate scan flip-flop} \label{App sff}
\begin{figure}[!htbp]
    \centering
    \includegraphics[scale=0.568]{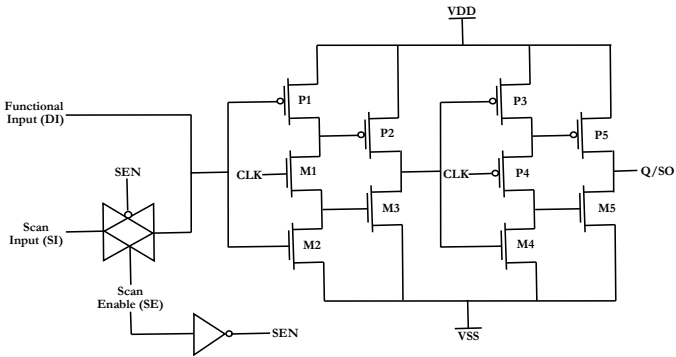}
    \caption{Proposed Approximate Scan flip-flop Schematic design}
    \label{fig:APPROXSFF}
\end{figure}

The functional path delay in the proposed approximate design is reduced by eliminating the multiplexer logic in the DI path. The DI input is directly connected to the flip-flop input, as shown in Fig. \ref{fig:APPROXSFF}. Due to the direct connection, it is observed that the DI path overdrives the SI path (SI is masked by DI so that the D flip-flop receives DI at its input irrespective of SI transitions) in some cases. This problem can be solved by increasing the drive strength of transistors in the SI path. We made aspect ratios of transistors constructing the transmission gate in the SI path as 2 $\times$ the aspect ratio of transistor M1 in Fig. \ref{fig:APPROXSFF} and verified the correct functionality.

In functional mode, the SE signal remains Low (logic `0'), so the functional input DI drives the scan flip-flop. Based on the logic value of DI, either P1 or M2 gets turned ON. The designs are negative edge triggered, so the transition at the output happens at the negative edge of the clock. Assuming DI is at a logic low during the positive edge of the clock, the transistors P1 and M1 are turned ON, and M2 gets turned off. Due to this, M3 turns on, and P2 turns off due to high logic at the gate terminals. $i.e.$ output at the master latch is logic `0'. The slave latch is not functional during the positive edge of the clock. Next, when the negative edge of the clock arrives, P4 and P3 get turned on, and M4 turns off. As the P3 transistor is ON, P5 turns off, and output Q is logic `0'. The opposite operation occurs when DI is at high logic.  The proposed design is negative edge triggered and has clock load only on two transistors.

In Test/scan mode, the SE signal remains at a logic high. As mentioned above, due to the direct connection of the functional input, sometimes it overrides the scan input. To overcome this, we made aspect ratios of transistors constructing the transmission gate in the SI path as 2 $\times$ the aspect ratio of transistor M1 in Fig. \ref{fig:APPROXSFF}. As the SE signal remains at logic high, the scan input (SI) passes to the scan flip-flop, and the same operation is carried based on the logic levels of the scan input (SI) and a clock signal. The approximate design gives less delay compared to the designs in Section \ref{Mux-based sff} \& \ref{GDI sff}. This design can be used to reduce the test time.

\subsection{Multiplexer-based scan flip-flop} \label{Mux-based sff}
\begin{figure}[!htbp]
    \centering
    \includegraphics[scale=0.608]{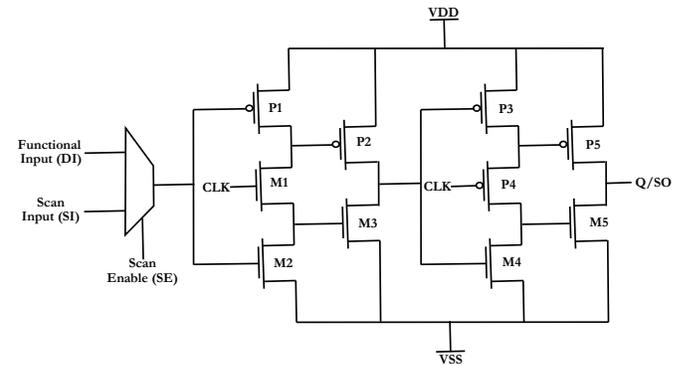}
    \caption{Proposed Mux-based Scan flip-flop Schematic design}
    \label{fig:MUXSFF}
\end{figure}

We adopt the negative edge triggered D flip-flop structure proposed in \cite{mux-based-sff} for the flip-flop and convert it into a scan flip-flop by adding a $2\times1$ multiplexer made of transmission gates. Fig. \ref{fig:MUXSFF} shows the schematic diagram of the proposed mux-based scan flip-flop. The proposed scan flip-flop uses a single clock signal to
enable/disable the master latch and the slave latch i.e., the clock load is present only on two transistors. A control signal named scan enable (SE) is used to select between the functional and test modes of operation. In the functional mode, SE=`0' and the output of the flip-flop follow the data input (DI) at the negative edge of the clock. During the test mode, SE=`1' and the output of the flip-flop follow the scan input (SI) at the negative edge of the clock. The functionality of the flip-flop is verified using Cadence Spectre with a 45-nm CMOS technology node. This scan flip-flop incurs a large delay due to the multiplexer present in the functional path. 

\subsection{Gate-Diffusion Input (GDI) based scan flip-flop} \label{GDI sff}
Gate-Diffusion Input (GDI) method \cite{GDI} allows for designing digital logic with low complexity, reduced power-delay product, and area. We use the GDI method to design the multiplexer part of the scan flip-flop described in Section \ref{Mux-based sff} by preserving the functionality. 

\begin{figure}[!htbp]
    \centering
    \includegraphics[scale=0.57]{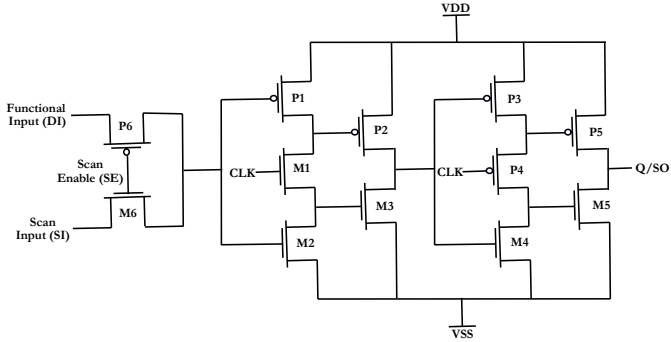}
    \caption{GDI-based Scan flip-flop Schematic design}
    \label{fig:GDISFF}
\end{figure}

Fig. \ref{fig:GDISFF} shows the schematic diagram of the proposed GDI-based scan flip-flop. In the functional mode (SE=`0'), the transistor P6 turns on and passes the functional input to the flip-flop. Whereas in the test mode (SE=`1'), the transistor M6 turns on and passes the scan input to the flip-flop. This design results in low power and area compared to the mux-based design reported in Section \ref{Mux-based sff}. The delay for this scan flip-flop is slightly higher than the delay for the scan flip-flop in Fig. \ref{fig:MUXSFF}. This is because the resistance of PMOS/NMOS is higher compared to the resistance offered by a transmission gate which is the parallel combination of resistances of PMOS and NMOS.

\section{Experimental Results and comparisons}
\begin{figure*}[!htbp]
    \centering
    \includegraphics[scale=0.2]{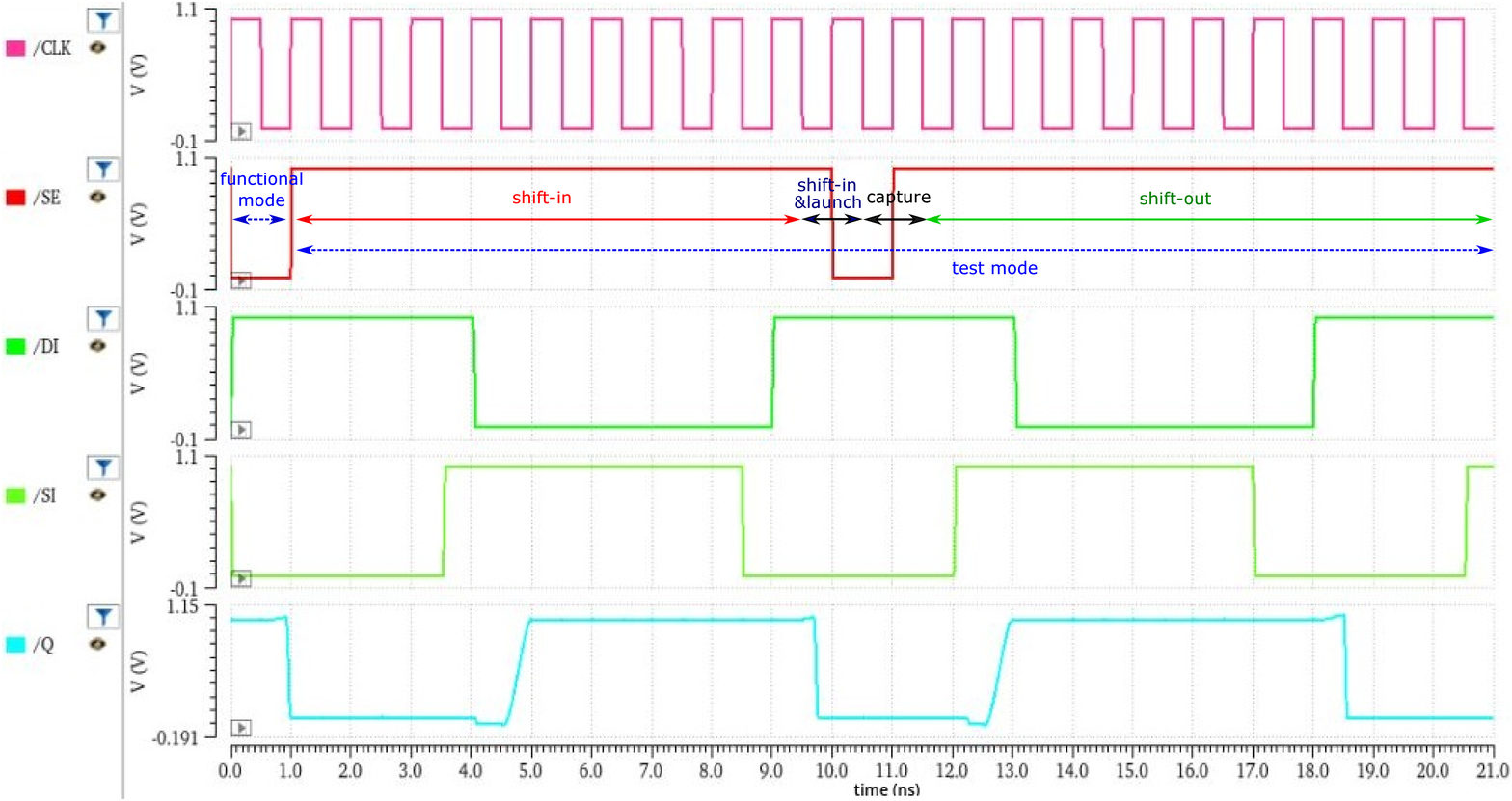}
    \caption{Post Layout Simulation result of the proposed Approximate scan flip-flop at 1GHz.}
    \label{fig:simresultsff1}
\end{figure*}

\begin{figure*}[!htbp]
    \centering
    \includegraphics[scale=0.2]{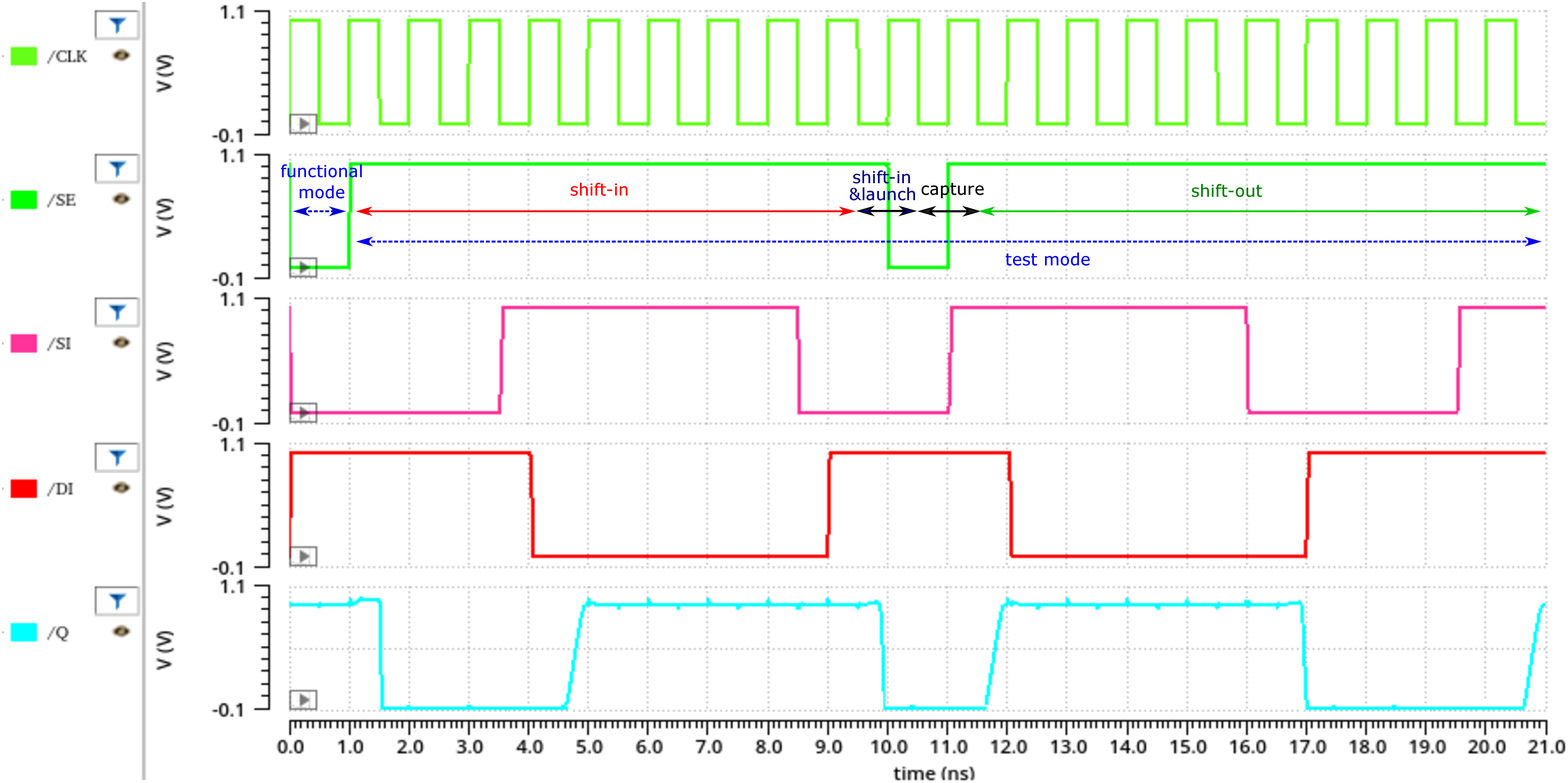}
    \caption{Post Layout Simulation result of the mux-based scan flip-flop at 1GHz.}
    \label{fig:simresultsff2}
\end{figure*}

\begin{table*}
\caption{Pre-layout delay and power for the proposed scan flip-flops in functional and test modes}
\begin{center}
\begin{tabular} {|c|c|c|c|c|c|c|c|c|c|c|} 
\hline 
\multirow{2}{*}{Design} & \multicolumn{4}{|c|}{Functional mode} & \multicolumn{4}{|c|}{Test mode}\\ \cline{2-9}
& $t_{su}$ (ns) & $t_{cq}$ (ns) & $t_{su}+t_{cq} = t_{pd}$ (ns) & Power ($\mu$W) & $t_{su}$ (ns) & $t_{cq}$ (ns) & $t_{su}+t_{cq} = t_{pd}$ (ns) & Power ($\mu$W)\\ 
\hline 
Mux-based & 0.058 & 0.141 & 0.19 & 2.65 &0.06 & 0.14 & 0.2 & 2.1 \\[1ex]
GDI-based & 0.18 & 0.14 & 0.32 & 0.56 &0.38 & 0.13 & 0.51 & 0.57\\[1ex]
Approximate & 0.06 & 0.14 & 0.2 &  0.41 & 0.04 & 0.14 & 0.18 & 0.44 \\
\hline
\end{tabular}
\label{tab:pre-layout}
\end{center}
\end{table*}
\begin{table*}
\caption{Post-layout delay and power for the proposed scan flip-flops in functional mode}
\begin{center}
\begin{tabular} {|c|c|c|c|c|c|c|c|} 
\hline 
Design & $t_{su}$ (ns) & $t_{cq}$ (ns) & $t_{su}+t_{cq} = t_{pd}$ (ns) & Power ($\mu$W) & Time gain w.r.t Mux-based & Power gain w.r.t Mux-based \\ 
\hline 
Mux-based & 0.088 & 0.283 & 0.371 & 3.62 & - & -\\[1ex]
GDI-based & 0.66 & 0.284 & 1.05 & 1.06 & -0.68ns & 70.7\%\\[1ex]
Approximate & 0.055 & 0.3 & 0.35 & 0.51 & 0.02ns & 85.9\%\\
\hline
\end{tabular}
\label{tab:simfunctional}
\end{center}
\end{table*}
\begin{table*}[!htbp]
\caption{Post-layout delay and power for the proposed scan flip-flops in test mode}
\begin{center}
\begin{tabular} {|c|c|c|c|c|c|c|} 
\hline 
Design & $t_{su}$ (ns) & $t_{cq}$ (ns) & $t_{su}+t_{cq} = t_{pd} $ (ns) & Power ($\mu$W) & Time gain w.r.t Mux-based & Power gain w.r.t Mux-based \\
\hline
Mux-based & 0.085  & 0.05 & 0.365 & 3.81 & - & -  \\[1ex]
GDI-based & 0.77 & 0.282 & 0.94 & 1.37 & -0.575ns & 64\% \\[1ex]
Approximate & 0.04 & 0.3 & 0.34 & 0.56 & 0.025ns & 85.3\%\\
\hline
\end{tabular}
\label{tab:simtest}
\end{center}
\end{table*}
The functionality of all flip-flops is verified by post-layout simulation using Cadence Spectre with a 45-nm CMOS technology node. The timing simulation was done up to 1GHz clock frequencies. The layout for all the proposed designs has been done using the Virtuoso physical layout suit. Minimum sizes were kept for transistors with a NMOS to PMOS W/L
ratio being 2, $i.e.$ $(W/L)_p$ = 2 $\times (W/L)_n$. We considered ten cycles each for shift-in and shift-out phases, assuming a scan chain consists of ten flip-flops. In general, a scan chain with n flip-flops requires n cycles each for shift-in and shift-out phases. A combinational circuit made of NAND gates drives the data input (the combinational part drives the data input as shown in Fig. \ref{fig:fullscan}). The proposed flip-flops are negative-edge triggered. We measured the setup time ($t_{su}$), clock-to-q propagation delay ($t_{cq}$) and average power for functional and test modes. During the functional mode setup time measurement, the clock is kept steady after the negative edge transition, scan enable is tied to `1’, and moved the scan input well before the clock negative edge transition. The scan flip-flop failed to follow the input when the input made a transition at some delay before the clock edge.  At this time, the time difference between the clock edge and the scan input edge gives the setup time. The same procedure is repeated in the test mode except that the scan enable signal is tied to `0’ and moves the data input instead of the scan input. The simulation results of the Approximate and mux-based scan flip-flop at 1GHz clock frequency are shown in Fig. \ref{fig:simresultsff1} and Fig. \ref{fig:simresultsff2} respectively. 

\begin{table}[!htbp]
\caption{Comparison of the proposed and existing scan flip-flop designs in terms of delay, power, and area in Test mode at 45nm}
\begin{center}
\begin{tabular} {|c|c|c|c|} 
\hline 
Design & $t_{su}+t_{cq} = t_{pd} $ (ns) & Power ($\mu$W) & Area\\
\hline
\cite{mishra2010modified} & 0.077 & NA & 26\\[1ex]
\cite{kumar2009robust} & 0.043   & 8.98  & 33  \\[1ex]
\cite{ahlawat2018high} & 0.674&NA & 38\\[1ex]
Mux-based & 0.36 & 3.81 & 16\\[1ex]
GDI-based & 0.94 & 1.37 & 12\\[1ex]
Approximate & 0.34 & 0.56 & 14\\
\hline
\end{tabular}
\label{tab:simcomparison}
\end{center}
\end{table}

The pre-layout simulation results for all the proposed flip-flops in functional and test modes are reported in Table \ref{tab:pre-layout}. Table \ref{tab:simfunctional} and \ref{tab:simtest}show the post layout timing simulation results of the proposed designs in functional and test mode. It is observed that the post-layout delay and power consumption are higher compared to the pre-layout results due to the added parasitic effects after layout. From Table \ref{tab:simfunctional} and \ref{tab:simtest}, the average power consumption is slightly higher in the test mode than in the functional mode for all flip-flop designs. This is due to the unnecessary combinational circuit transitions and increased switching activity in the flip-flops caused by a serial shifting of test vectors in scan-based design. The delay is high for the GDI-based design that uses a multiplexer made of pass transistors in the functional path. Whereas the delay for the approximate design is less due to the absence of multiplexer logic in the functional path. The time gain for the approximate design with respect to the multiplexer-based design is 0.02ns in the functional mode and 0.025ns in the test mode.  The power gain for the GDI-based design with respect to the multiplexer-based design is 70.7\% in the functional mode and 64\% in the test mode. And the power gain for the approximate design with respect to the multiplexer-based design is 85.9\% in the functional mode and 85.3\% in the test mode. Among the three designs, it is observed from Table \ref{tab:simfunctional} and \ref{tab:simtest} that the approximate design is efficient in terms of delay and power consumption.


The comparison of test mode post-layout simulation timing and average power consumption results with the existing scan flip-flops are reported in Table \ref{tab:simcomparison}. For a fair comparison, the results for all the scan flip-flops should be calculated with the same technology node. We calculate the delay and average power values for the existing scan flip-flops at 45nm technology. To do this, We used an
easy platform, "DeepScaleTool," proposed \cite{sarangi2021deepscaletool} to obtain reliable scaling factors for different
design parameters in the deep-submicron era. It is a spreadsheet-based tool where the scaling factors can be obtained by entering the current and target technology nodes. The target node parameters can be obtained by dividing the current node parameters by the scaling factors. From Table \ref{tab:simcomparison}, the delay is high for the GDI-based design that uses a multiplexer made of pass transistors in the functional path. The power consumption for all three proposed scan flip-flop designs is less compared to the existing scan flip-flop designs. Moreover, all the proposed designs require less area than the area reported in the existing works. 

\section{Conclusion}
In this work, we introduced three different scan flip-flops: mux-based, GDI-based, and approximate designs with low average power consumption and less area. The delay introduced by the multiplexer in the functional path is eliminated in the proposed approximate design resulting in less delay. All three designs give less power consumption compared to the existing scan flip-flop designs.  The proposed Approximate scan flip-flop can significantly reduce the test power and time when used in scan chains of large lengths. 

\bibliographystyle{IEEEtran}
\bibliography{references}

\end{document}